\documentclass[useAMS,usenatbib]{mn2e}
\usepackage{rotating}
\usepackage{journals}
\usepackage{graphicx, subfigure}
\usepackage{xspace}
\usepackage{amssymb}
\usepackage{mathtools}
\usepackage[T1]{fontenc}
\usepackage{ae,aecompl}

\LARGE \normalsize \title[{VLT spectroscopy of J1357.2 in Quiescence}]{VLT spectroscopy of the Black Hole Candidate  Swift J1357.2-0933 in Quiescence}
\author[Torres et al.]  {M.A.P.~Torres$^{1,2,3}$\thanks{email : M.Torres@sron.nl}, P.G.~Jonker$^{1,2}$,  J. C. A. Miller-Jones$^{4}$,  D.~Steeghs$^{5}$,  
S. Repetto$^{2}$,  
\newauthor J. Wu$^{6}$ \\
$^1$SRON, Netherlands Institute for Space Research, Sorbonnelaan 2, 3584~CA, Utrecht, The Netherlands\\
$^2$Department of Astrophysics/ IMAPP, Radboud University Nijmegen, Heyendaalseweg 135,6525 AJ, Nijmegen, The Netherlands \\
$^3$European Southern Observatory. Alonso de C\'ordova 3107, Vitacura, Casilla 19001, Santiago de Chile, Chile \\
$^4$International Centre for Radio Astronomy Research, Curtin University, GPO Box U1987, Perth, WA 6845, Australia \\
$^5$Department of Physics, University of Warwick, Coventry CV4 7AL, UK
\\
$^6$Harvard--Smithsonian Center for Astrophysics, 60 Garden Street, Cambridge, MA~02138, U.S.A.\\
}

\begin{document}

\maketitle

\begin{abstract} \noindent 

We present time-resolved optical spectroscopy of the counterpart to
the high-inclination black hole  low-mass X-ray binary Swift
J1357.2-0933 in  quiescence.  Absorption features from the mass donor
star were not detected.  Instead the spectra display prominent broad
double-peaked H$\alpha$ emission  and weaker He{\sc i} emission
lines. From the H$\alpha$ peak-to-peak separation we constrain the
radial velocity  semi-amplitude of the donor star to ${K_2} >
789$ km s$^{-1}$. Further analysis through radial velocity and
equivalent width measurements indicates that the H$\alpha$ line is
free of variability  due to  S-wave components or disc eclipses. From
our data and  previous observations during outburst, we conclude that
long-term  radial velocity changes ascribed to a precessing disc were
of low amplitude or not present.  This implies that  the  centroid
position of the line should closely represent the systemic radial
velocity, $\gamma$. Using the derived $\gamma = -150$ km
  s$^{-1}$ and the best available limits on the source distance, we
  infer that the black hole is moving towards the Plane
in its current Galactic orbit unless the proper motion is substantial. 
Finally,  the depth of the central absorption in the
double peaked profiles  adds support for Swift J1357.2-0933 as a
high-inclination system. On the other hand, we argue that the low
hydrogen column density inferred from X-ray fitting suggests that the
system is not seen edge-on.

\end{abstract}


\section{Introduction}

The X-ray transient Swift J1357.2-0933 (hereafter J1357.2) was
discovered in outburst on 28 Jan 2011 with the Burst Alert Telescope
on-board {\it Swift}  (Krimm et al.  2011a).  Follow-up
observations with the {\it Swift} X-ray Telescope  revealed a
spectrum consistent with an absorbed power law with a low
hydrogen column density $N_H$ of $(1.2 \pm 0.7) \times 10^{20}$
cm$^{-2}$ and a photon index that evolved from $\Gamma = 1.5$ to 2.1
during the outburst decline (Krimm et al.  2011b, Armas Padilla et
al. 2013).  A 0.5-10 keV outburst peak luminosity of $1.1 \times 10^{35} \times
(\frac{d}{1.5~kpc})^2$ erg s$^{-1}$ is inferred from these
observations.  In addition,  XMM-Newton data obtained on 5 Feb 2011
could be fitted with a three component model: the above mentioned power
law with a 93 percent contribution to the total flux, a thermal (disc)
component with $kT = 0.2$ keV, and one edge at 0.73 keV. The thermal
component is interpreted as soft emission originating in the accretion
disc, while the edge is associated to interstellar iron.  
The low X-ray peak luminosity together with  the
evolution (softening) of the power-law dominated spectrum, the low
temperature for the thermal component associated to the accretion disc
and the detection of radio emission  (Sivakoff, Miller-Jones \& Krimm
2011)  are in accordance with the source being a black hole system in
the low-hard state through the entire outburst. In this regard, timing
analysis of both the XMM-Newton data as well as contemporaneous RXTE
Proportional Counter Array observations showed a power spectrum
with characteristics more similar to those detected in the low-hard
state of black holes than neutron stars (Armas Padilla et al. 2014a).
On the other hand, early in the outburst a 6 mHz quasi-periodic
oscillation was detected in one of the RXTE observations. This low
frequency is atypical in black holes and more common in dipping
neutron stars (see Armas Padilla et al. 2014a for details).

J1357.2 was observed in quiescence  with XMM-Newton on 10 July 2013
and detected at a 0.5 - 10 keV X-ray luminosity of $8.5^{+5.5}_{-2.6}
\times 10^{29} \times (\frac{d}{1.5~kpc})^2$ erg s$^{-1}$ (Armas
Padilla et al. 2014b).  The X-ray spectrum in quiescence is
consistent with an absorbed power law with $\Gamma = 2.1 \pm 0.4$,
although fits to single thermal models were also permitted  by the
data. 

The optical and near-infrared counterparts to J1357.2 were identified
with a $r'\simeq 16.30$ and $K = 17.4$ mag star\footnote{The
  coordinates in Rau et al. (2011) are incorrect due to a
  typographical error. The coordinates for J1357.2 are $\alpha(J2000)$ =
  13:57:16.829 and $\delta(J2000)$= -09:32:38.75 as provided by SDSS} (Rau et
al. 2011). A high-resolution optical spectrum obtained on 2 February
2011 showed no clear indication for emission lines, lack of diffuse
interstellar bands and the presence of weak absorption lines from the
interstellar Na doublet (Torres et al. 2011). A low-resolution
spectrum taken one day later exhibited weak and broad H$\alpha$ and
H$\beta$ emission lines with the former having 7 \AA~equivalent width
(EW) and $\sim4000$ km s$^{-1}$ Full Width at Zero Intensity (FWZI;
Milisavljevic et al. 2011).  High-resolution H$\alpha$ spectroscopy
undertaken on 25-27 February 2011 resolved the very broad
double-peaked line profile with a peak-to-peak separation of
$\sim1800$ km s$^{-1}$ (Casares et al. 2011).  Further time-resolved
observations allowed Corral-Santana et al. (2013) to find a $2.8 \pm
0.3$ hr modulation in radial velocities extracted from the H$\alpha$
line profile. Assuming this periodicity corresponds to the orbital
period and using the line peak-to-peak separation to constrain the
radial velocity semi-amplitude of the donor star to ${K_2} \geq 690$
km s$^{-1}$, a mass larger than $3.0$ M$_\odot$ is estimated for the
compact object. Complementary time-resolved optical photometry
did not show any evidence for variability with the above periodicity
(Corral-Santana et al.~2013). The light curves show dipping
variability with a recurrence time that gradually increases from 2.3
min to 7.5 min over 69 days of follow-up photometry.  The recurrent
optical dips (up to 0.8 mag) are interpreted as obscuration events
caused by a toroidal structure with asymmetric height with respect to the
orbital plane.  In this scenario, the changes in the dipping frequency
would imply that this structure is moving inside-out the disc during
outburst. In order to produce the profound dips in the light curve, a
very large system inclination is necessary.  However, neither the
optical data nor the X-ray data showed dips or eclipses due to the
donor star (Armas-Padilla 2013a, Corral-Santana et al.~2013). To explain the
non detection of these features in the light curve, Corral-Santana et
al.~(2013) invoke an accreting binary configuration in which the mass ratio
is low enough to have a Roche-lobe filling donor star with a radius
smaller or similar to the disc outer rim.

The optical counterpart to J1357.2 was also identified in quiescence
at $r′ = 21.96$ in Sloan Digital Sky Survey (SDSS) images taken in May
2006 (Rau et al. 2011).  Based on the SDSS pre-outburst colors and
assuming no disc contribution to the optical light, the donor star was
tentatively classified as an M4 dwarf at $\sim1.5$ kpc distance.  This
is in good agreement with the M4.5 donor at $\sim1.6$ kpc expected
from a Roche-lobe filling donor in a 2.8 hr orbital period system
(Corral-Santana et al. 2013).  However, time-resolved photometry
during quiescence at different post-outburst epochs (Shahbaz et
al. 2013) reveals a very different picture: the donor star is not the
dominant source of light at optical or infrared wavelengths. The
multicolor light curves lack evidence of any orbital variability
caused by the donor star, i.e.~ellipsoidal modulations, eclipses or
dip features. Instead the photometry shows flares of up to $\sim1.5$
mag and $\sim2$ mag at optical and infrared wavelengths,
respectively. This is the largest flare amplitude known in quiescent
black hole X-ray binaries.  Moreover, the dipping behaviour observed
in outburst persists in quiescence but with a $\sim30$ min recurrence
time. Both the optical to mid-infrared quiescent spectral energy
distribution (SED) and optical variability SED of J1357.2 can be
described by a single power-law model with index $\Gamma \sim 1.4
$. This steep power-law together with the high amplitude flickering
are interpreted as due to synchrotron emission from a variable, weak
jet that dominates the SED over the donor star light (Shahbaz et
al. 2013). Using the relation between orbital period and the outburst
optical amplitude, Shahbaz et al.  constrain the distance towards
J1357.2 to be between 0.5 and 6.3 kpc.

The lack of accurate dynamical parameters and observational details of
the  accretion disc structure during quiescence motivated us to
perform time-resolved optical spectroscopy.  We begin with a
description of the observations and data reduction steps (Section
2). The average optical  spectrum of J1357.2 in quiescence is
presented in Section 3 where we also characterize the H$\alpha$
line profile and constrain  its variability. Finally, the results are
presented and discussed in Section 4, where we examine key questions
such as the orbital parameters and space velocity for the
  system. Our conclusions are summarized in Section 5.  

\section{Observations and data reduction}

Time-resolved optical spectroscopy of J1357.2 was obtained with the
FOcal Reducer and low dispersion Spectograph 2 (FORS2, Appenzeller et
al. 1998) which is mounted on the Cassegrain focus of the 8.2-m ESO
Unit 1 Very Large Telescope at Paranal, Chile.  The observations were
obtained in service mode under program 091.D-0865(A) during 14 April
2013 5:37 - 6:22 UT, 18 April 2013 4:35 - 8:00 UT and 4 May 2013 4:12
- 4:57 UT.  FORS2 was used with the standard resolution collimator and
the $2048 \times 4096$ pixels MIT two CCDs mosaic detector binned
$2 \times 2$ to provide a 0''.25 pixel$^{-1}$ scale. The instrument
was operated in long-slit mode with the 600 line mm$^{-1}$ grism
GRIS\_600RI.  J1357.2 was centered in a 1''.0 wide slit with its
location offset from the CCD center. This instrumental setup yields a
dispersion of 1.6 \AA~ pixel$^{-1}$, a coverage in the spectral range
$\lambda\lambda5300-8630$ and a slitwidth-limited resolution of
$\sim5$ \AA~ full-width half maximum (FWHM).  A total of six one
hour-long observing blocks (OBs) were executed consisting of four
spectroscopic integrations of 640 s each.  Four, sixteen and four
spectra were collected on 14 April at $\sec~z < 1.08$, 18 April from
$\sec~z$ 1.04 to 1.43 and 4 May at $\sec~z < 1.07$, respectively. We
measure from the width of the source spatial profile at spectral
positions covering H$\alpha$ an image quality between
$0''.6 - 0''.8$, $0''.5 - 0''.7$ and $0'.'7 - 0''.8$ for the first,
second and third nights, respectively and a mean $0''.66 \pm 0''.07$
FWHM from the three nights.  Therefore the observations were obtained
in seeing-limited conditions yielding a spectral resolution of $\sim3$
\AA~ FWHM corresponding to 140 km s$^{-1}$ at H$\alpha$.

The spectra were reduced and extracted using standard techniques
implemented in the {\sc starlink}, {\sc figaro}, and {\sc pamela}
packages while the wavelength calibration was done with {\sc molly}.
The data reduction consisted of de-biasing and flat-fielding the
data. The spectra were extracted using the algorithm of Horne (1986)
to optimize the signal-to-noise ratio of the resulting
spectra. Exposures using comparison arc lamps were performed after the
end of each night in order to establish the pixel-to-wavelength
scale. This was derived through polynomial fits to 27 arc lines. The
rms scatter of the fit was always $< 0.07$ \AA, which is less than
$1/22$ of the wavelength dispersion.  The sky [O{\sc i}]
$\lambda\lambda 5577.34,6300.3$ lines and the OH emission blend at
7316.3 \AA~showed that wavelength zeropoint disparities reached up to
$\lesssim 30$ km s$^{-1}$ in amplitude. These deviations were
corrected for by applying to the spectra zeropoint shifts calculated
using the [O{\sc i}] $\lambda6300.3$ emission feature.  The
resulting 24 spectra were normalized by dividing each of them by a low-order
spline fit to the continuum after masking out emission lines and
atmospheric absorption bands. Finally, the resulting spectra were
rebinned to a uniform pixel scale.

Since all the observations were seeing limited, systematic effects due
to excursions of the target position with respect to the slit centre
may affect the radial velocity determinations presented in this
  work.  In this regard, J1357.2 was centered on the slit at
the start of each OB execution, except perhaps during the last
two OBs on 18 April. For these OBs we lack the re-acquisition and
  through-slit images commonly saved when  the centering is checked or
  performed during service mode observations. In what follows we will conservatively assume that for
  both OBs the centering step was skipped. Hence, positional
  departures from the exact slit center and changes in the target
  position during the spectroscopic exposures have certainly
  happened. Unfortunately we lack the data required to quantify and
  correct for this effect (see e. g. Bassa et al. 2006). Section
2.4.3 in the ESO FORS2 manual (Issue 92.0) offers constraints on
the image motion for a given $z$ due to instrument
flexures. This motion increases with $z$.   As our observations were
performed after culmination, $z$ increased monotonously and it is
therefore reasonable to assume that the offsets in the target position
accumulate. Thus, we estimate a cumulative offset $<0.3$ (binned)
pixels at the end of  the execution of the single OBs during 14 April
and 4 May.  The cumulative offsets at the end of the four OBs executed
on 18 April are estimated to be $<0.3, <0.3, <0.7, <1.1$ binned
pixels. The systematic effects could therefore have introduced  radial
velocity offsets (at H$\alpha$) from $<22$ to $<84$ km s$^{-1}$. 
  We have measured the positional offsets along the spatial direction.
  For this we have calculated the centroid for the spatial profile of
  J1357.2 at H$\alpha$ in all spectra. The centroids differ by < 0.3
  (binned)  pixels.  This suggests that the cumulative positional
  offsets in the dispersion direction during 18 April may be smaller
  than estimated if they have a similar amplitude to that measured 
for the offsets in the spatial direction.

\section{Data analysis}
\subsection{Averaged spectral features}

In Fig. 1, we present the result obtained by averaging the 24
individual FORS2 spectra. The averaged data show no evidence for
photospheric features from the donor star such as the TiO bands
characteristic of M-type stars. The interstellar sodium doublet
  found in outburst by Torres et al. (2011) is not obvious due to the
low reddening towards the source and the fact that this feature falls
on top of the He{\sc i} $\lambda5876$ emission line.  All significant
absorption features in the spectrum are caused by the Earth's
atmosphere.  The spectrum is dominated by a prominent broad and
double-peaked H$\alpha$ emission line together with weaker emission
lines of He{\sc i} $\lambda\lambda5876,6678,7065$.  He{\sc i}
$\lambda5876$ is also double-peaked while He{\sc i} $\lambda6678$ is
partially resolved from the H$\alpha$ line red wing.  At red
wavelengths there is no obvious emission from Paschen lines or the two
Ca{\sc ii} triplet components covered by the data. The apparent
emission feature at $\sim\lambda8492-3$ is not coincident with any of
the above Hydrogen or Ca{\sc ii} lines and it is most likely an
artifact.

The mean H$\alpha$ and He{\sc i} $\lambda5876$ double-peaked line
profiles  were fitted with  1 and 2-Gaussian functions using the
Marquardt algorithm (Bevington 1969). From the 1-Gaussian fit we
derive the line FWHM given in Table 1. The  2-Gaussian model allows us
to measure the velocity shifts of the blue ($V_b$) and red ($V_r$)
peaks with respect to the line rest wavelength.  Their difference ${\Delta V^{pp}}
= {V_r} - {V_b}$ (peak-to-peak separation) and mean $({V_r} +
{V_b})/2$ (centroid of the line) are included in Table 1.  This Table
also lists the measured full-width at zero intensity (FWZI) and
equivalent widths (EWs) for H$\alpha$ and He{\sc i}
$\lambda\lambda5876,7065$. The mean FWHM  for H$\alpha$ is
  consistent with the $\sim3900$ km s$^{-1}$ FWHM  found for this line
  in low-resolution spectroscopy obtained on 29 April 2013 UT (Shahbaz
  et al. 2013), although on that occasion the line profile appeared
  single peaked and with $-120$ \AA~EW. This value is   below the
  maximum EW observed when studying the time variability of
  H$\alpha$ (Section 3.2).

  The mean peak-to-peak velocity separations for the H$\alpha$ and
  He{\sc i} $\lambda5876$ lines are $2340 \pm 20$ km s$^{-1}$ and
  $2640 \pm 70$ km s$^{-1}$, respectively. This difference suggests
  that the He{\sc i} emission line originates from regions in the disc
  closer to the black hole than the regions responsible for the
  H$\alpha$ emission. The FWHM and velocity separation of the double
  peaks in the H$\alpha$ line (Table 1) are a factor 1.2-1.3 larger
  than the values reported during outburst ($\sim3300$ km s$^{-1}$
  FWHM and ${\Delta V^{pp}} =1790 \pm 67$ km s$^{-1}$, Corral-Santana
  et al. 2013).  This difference is expected for a disc with a
  Keplerian velocity field since during outburst the disc increases
  its radius, thereby decreasing the velocity of the outer disc
  regions.The $\sim138$ \AA~ FWZI of the H$\alpha$ and He{\sc i}
  $\lambda5876$ implies projected velocities $\gtrsim 3150$ and
  $\gtrsim 3520$ km s$^{-1}$ for the inner part of the accretion disc
  emitting at these wavelengths.  The FWZI found for H$\alpha$
  is a factor $\sim1.6$ larger than observed at different times during
  the outburst (Milisavljevic et al. 2011, see also fig. 1 in
  Corral-Santana et al. 2013). This difference in FWZI can be
  explaned if the luminosity of the continuum decreased quicker than
  the luminosity of the emission lines when the source went from
  outburst to quiescence.

\subsection{H$\alpha$ line profile time variability}

We studied the variability
of the H$\alpha$ line with diverse methods. These included single and
2-Gaussian model fitting, the
double-Gaussian technique described by Shafter et al. (1986) and line
EW measurements. The results obtained from these techniques are shown
in Fig. 2.  A single Gaussian fit to the line profiles yields radial
velocities ranging from -189 to +118 km s$^{-1}$.  These radial velocities
reflect variations in the peak-to-peak intensities rather than changes
in the line centroid (see below).  The change in the double-peak
intensity from almost symmetric peaks to enhanced red-shifted or
blue-shifted peaks is obvious in Fig. 3 where we show the 
line profiles.  Even though this behavior
could be naively attributable to an S-wave originating in a hot spot
and/or the donor star, further inspection of the individual data in
Fig. 3 shows that the presence of such a narrow component is not
obvious.  In addition, these variations in the profile structure can
also not be explained as due to the eclipse of the accretion disc by
the donor star.  In such a scenario the blue-shifted disc emission is
first eclipsed while the red-shifted emission is eclipsed afterwards
since the donor star initially occults disc regions moving towards the
observer and subsequently blocks the receding regions from view.  This
is a distinctive rotational disturbance of the line profiles during
eclipse, the so-called Z-wave. Thus the changes in the H$\alpha$ line
profile of J1357.2 do not show the Z-wave characteristic of emission
line eclipses.  The variations in the peak intensities are likely due
to a highly variable non-uniform disc brightness distribution or/and
departures from an axisymmetric flat disc.  In this regard, we observe
changes of up to 10 percent in the peak-to-peak separation over
successive spectra (see below). The short-term (sub-orbital) changes
in the H$\alpha$ line profile morphology are reminiscent of those
observed in the black hole low-mass X-ray binary (LMXB) GRO J0422+32 (${P_{orb}} = 5.1$ hr,
q=0.12) in quiescence (see fig. 1 in Filippenko et al. 1995).  Another
variable feature in the line shape is the depth of the central
absorption delimited by the two line peaks.  While the averaged line
profile shows a moderately deep absorption, the individual spectra
reveal that its depth can vary significantly during consecutive
spectra. As observed in the first night (top-left panel Fig.3), the
absorption core passes from being near the continuum level in the
first spectrum to well above the continuum in the next one. Such
deep cores are expected from discs observed at high inclination (see
discussion).
 
Radial velocities were also derived with the double-Gaussian technique
(Shafter et al. 1986) which consists of convolving the line profiles with two Gaussian
bandpasses with separation $a$.  After  several trials, we chose to
use  a 200 km s$^{-1}$ FWHM for both Gaussians. The separation $a$
between the Gaussians was varied from 2300 km s$^{-1}$ to 6000 km
s$^{-1}$ in steps of 100 km s$^{-1}$.  The radial velocity  curves
obtained with Gaussian separations $a\sim2700 - 3200$  km s$^{-1}$
displayed a modulation with a higher amplitude than  found in the radial velocities
determined with a 1-Gaussian fit. We display in Fig. 2 the results for a Gaussian
separation of 2800 km s$^{-1}$.  At Gaussian separations larger than
$3300$  km s$^{-1}$, the radial velocities obtained with the
double-Gaussian technique show no obvious modulation. Ideally
  the wings of  disc emission lines should trace the motion of the
  compact object. In fact, by  applying the above technique to their
  outburst data,  Corral-Santana et al. (2013) recovered
  radial velocity modulations beyond $a=3300$ km s$^{-1}$ - note here
  that the H$\alpha$ line  profile was narrower during outburst.   In
  this way they were able  to  estimate the orbital period and
  the radial velocity semi-amplitude of the primary (${K_1} =
43 \pm 2$ km s$^{-1}$)  for J1357.2.  Our  failure to do this could be primarily due to the
  systematic effects affecting the radial velocities (Section 2). The systematic
velocity shifts could be comparable to or larger than $K_1$. Additionally, systematic effects
are expected due to contamination of the red line wing by the overlapping He{\sc i} $\lambda6678$ emission
  line. 

The 2-Gaussian model was employed to calculate both the peak-to-peak
  separation and the line centroid.  The changes in the line
  morphology from near-symmetric to asymmetric double-peaked velocity
  profiles were taken into consideration when calculating these
  parameters. This was done because the
Gaussian component accounting for the weaker peak in the line will fit
also part of the broader base of the profile. This yields a velocity
that could be significantly offset from the real peak position in
question. To avoid this possible bias effect we examined each
individual fit and rejected from our analysis $V_b$ and $V_r$ when
either of the related Gaussian components had $FWHM > 45$ km s$^{-1}$.
This limit proved to be an effective quantitative way to select
strongly (eight in total) asymmetric profiles - for comparison the
FWHMs of the 2-Gaussians fit to the mean profile in section 3.1 is
$\sim35$ km s$^{-1}$.  Note also that this non-physical two component
model yields reduced ${{\chi}^2}=1.0$ to 1.4, providing therefore
statistically sound fits.  By using profiles for which both Gaussian 
components have $FWHM < 45$ km s$^{-1}$ (16 spectra), we derive a mean
peak-to-peak separation of $\overline{{V_r} - {V_b}} = 2370 \pm 70$ km
s$^{-1}$ and centroid of $(\overline{{V_r} + {V_b}})/2 = -130 \pm 60$
km s$^{-1}$. The uncertainties here and below correspond to the rms
scatter.  We also calculated the above line parameters by 
including  velocity shifts obtained  from the asymmetric profiles.
For these profiles only one of the 2-Gaussian components in the fit is narrow ($< 45$ km s$^{-1}$
FWHM) yielding $V_b$ (7 spectra) or $V_r$ (1 spectrum). 
 Taking the mean of all the reliable velocity shifts ($\overline{V_b}$, $\overline{V_r}$), we 
derive $\overline{V_r} - \overline{V_b} = ({1060 \pm 50}) - ({-1325 \pm
  80})= 2390 \pm 100$ km s$^{-1}$ and ${\overline{V_r} +
  \overline{V_b}})/2 = -130 \pm 50$ km s$^{-1}$. The results from both
calculations are fully consistent.  The temporal
variability for these two line parameters derived from the 2-Gaussian model fitting
is also displayed in Fig. 2 (top and middle panels).   On 18 April the line centroid shows no clear modulation with
velocities ranging from -222 km s$^{-1}$ to -50 km s$^{-1}$. The
amplitude of the variations is lower than that obtained from
velocities derived with the 1-Gaussian model (300 km s$^{-1}$ amplitude)
and double-Gaussian technique (650 km s$^{-1}$). 
On 4 May the line centroid is
constant during the 45 min length of the observations with a mean
value of $-180 \pm 6$ km s$^{-1}$.  During the three nights the
peak-to-peak separation ranges from 2200 to 2530 km s$^{-1}$.
Variations in the peak-to-peak separation can occur in short intervals
as observed on 14 April and 5 May when in a $\sim20$ min interval
changes of $130 \pm 50 $ and $200 \pm 60$ km s$^{-1}$ in amplitude
take place.  The fast changes (sub-orbital) in the peak-to-peak
separation and intensities indicate that there are azimuthal
variations in the outer disc velocities or structures.

The H$\alpha$ EW varies irregularly as shown in the bottom of
Fig. 2. With our 640s integration, the EW ranges values from $-73$ to $-129$
\AA~with a mean of $-100 \pm 14$ \AA.  The 14 April EW curve lacks
any evidence for a decrease or increase in the line EW due to eclipses
of the disc by the donor star.  A decrease in the EW would occur at
and near mid-eclipse if the disc regions responsible for the continuum
and line are obscured from view by the donor star.  On the other hand,
an increase in EW can be associated with emission line regions in the
disc that are not tightly confined to the orbital plane and thereby
visible during eclipse. Examples of the increase in the H$\alpha$ EW
at eclipse time can be found in the neutron star LMXB sources X1822-371
(${P_{orb}} = 5.1$ hr; Harlaftis et al. 1982) and EXO 0748-676
(${P_{orb}} = 3.8$ hr; Pearson et al. 2006).  None of these
characteristics are present in the EW curve which appears dominated by
erratic variations. For comparison with other black hole LMXBs in
quiescence, the strength of H$\alpha$ is not above than that found in GRO
J0422+32 ($-181$ to $-214$ \AA; Harlaftis et al. 1999).

\subsection{Further analysis: cross-correlation and H$\alpha$ Doppler tomography}

In order to search for signatures of the donor star that could have
been canceled out by averaging the data, we cross-correlated the
individual spectra against SDSS stellar templates covering spectral
types between F and M. The cross-correlation was performed over
sections free of telluric and emission line features.  No
cross-correlation functions were discernable from this analysis.

In an attempt to corroborate the 2.8 hr orbital period, we performed
Doppler tomography on the H$\alpha$ line by phase folding the data
over different trial periods and adopting a systemic radial velocity of -150
km s$^{-1}$ (see Section 4.2). The tomograms (not shown) lead to
variable asymmetries in the disc emission, but not to the extent that
there was a clear favoured period.  The reason why we were unable to
use the tomographic reconstruction to constrain the orbital period is
the lack of an S-wave that would have resulted in a strong emission
feature in the tomogram.

\section{Discussion}

Our optical spectroscopy of J1357.2 during quiescence shows no
evidence for absorption features from the donor star.  We therefore
make use  of the strong double-peaked H$\alpha$ emission line  to set
constraints on the system parameters. For this, comparison will be
drawn between the spectroscopic and photometric properties of J1357.2
and those observed in relevant types of accreting binaries such as
accretion disc corona (ADC) systems and ultra-compact binaries
with a white dwarf accretor  (better known as AM CVn systems). ADC sources are
persistent LMXBs with an orbital inclination of $>80^{\circ}$. AM CVn
systems  have in common with short orbital period black hole LMXBs a
low mass ratio $q={M_2}/{M_1} \lesssim 0.1 $, where $M_2$ and $M_1$
are the mass of the donor star and accreting object, respectively. 

\subsection{The radial velocity semi-amplitude of the donor star}

Assuming an axisymmetric disc with gas in Keplerian motion,  the
projected motion of the outer edge of the disc $V_d$ can be used to
constrain the radial velocity semi-amplitude of the donor star ($K_2$)
given that ${K_2} < {V_d}$. $V_d$ has been determined for eight
  black hole LMXBs in quiescence  by modelling the phase averaged
  H$\alpha$ line profile and by locating the hotspot in Doppler
  tomograms (e.g. Orosz et al. 1994 and Marsh et al. 1994). In this
  way it has been found that ${V_d}(H\alpha)/{K_2}$ is between 1.11
  and 1.47 (table 2 in Orosz et al. 2002).  Using this empirical relationship and the H$\alpha$ line
  observed in outburst, Corral-Santana et al. (2013) constrained
  ${K_2} > 716 \pm 26$ km s$^{-1}$. For this they adopted
    ${V_d}(H\alpha)/{K_2}=1.25$ (Orosz et al. 1994, Orosz \& Bailyn 1995) and took into
    account  that, since the source is in outburst, the peak-to-peak
    separation is less than $ 2 \times {V_d}(H\alpha)$ during quiescence.  Following a
    similar approach, we constrain $K_2$ to be $\gtrsim 796 \pm 7$ km
  s$^{-1}$ ($K_2 > 789$ km  s$^{-1}$) by conservatively using  ${V_d}(H\alpha)/{K_2} \leq 1.47$ during quiescence, the observational fact that
   $2 \times {V_d}(H\alpha) \gtrsim {\Delta V^{pp}}$ and ${\Delta V^{pp}}=2340 \pm 20$
  km s$^{-1}$ as found from the averaged H$\alpha$ profile 
  during quiescence (Section 3.1). This implies a mass function
  $f(M{_1}) > 6.0$ M$_\odot$ if J1357.2 is in a $2.8 \pm 0.3$ hr orbit.  

Adopting the radial velocity semi-amplitude of the primary (${K_1} =
43 \pm 2$ km s$^{-1}$; Corral-Santana et al. 2013) derived by applying
the double-Gaussian technique to outburst data of the H$\alpha$
emision line, a constraint for the mass ratio $q = {K_1}/{K_2}
\lesssim 0.054$ is obtained.

\subsection{The systemic radial velocity}

The characterization of the H$\alpha$ profile performed in Section 3.2
shows that its centroid is offset to the blue on average by $130 \pm 50$ km
s$^{-1}$.  The position of a line originating in a disc
will be shifted from its rest wavength by the systemic radial velocity and primary's radial
velocities at the time of the observation as: $\gamma + {K_1}\sin 2 \pi
\phi$, where $\gamma$ is the systemic radial velocity and $\phi$  is the
orbital phase. Additional velocity components are usually present. They
are due, for instance, to emission from the gas stream, hot spot and/or
donor star. Furthermore, departures from a disc with gas in Keplerian
motion are possible in LMXBs due to tidal effects on
the disc. The tidal action of the donor on the outer parts of the disc
can excite spiral shock patterns or cause the disc to elongate and
precess. Disc precession can dominate the orbital-averaged velocity
shifts observed in the emission line centroid.  These contributions to
the velocity profile can make it difficult to reliably  establish $\gamma$ or
${K_1}$ from time-resolved spectroscopy of  emission lines (see e.g. Orosz 
et al. 1994 and sec 7.5 in Shahbaz et et al. 2013).  

In what follows we present and discuss three pieces of observational evidence
that indicate that the $-130$ km s$^{-1}$ offset of the H$\alpha$ line
centroid in J1357.2 most likely represents (within the errors) the systemic radial velocity: 
first, as reported in Section 3.2. the double-peaked
H$\alpha$ line profiles lack the presence of any narrow line
component with velocity position modulated with the orbital motion.
Therefore, J1357.2 shows no S-wave associated to emission from
confined regions in the accretion flow (as discussed above) that can
produce a radial velocity offset in an averaged line profile when the
orbital phase coverage during the observations is non-uniform. S-waves
are features commonly present in the time-resolved spectra of
accreting binaries, but exceptions exist where this feature is not
detected or is very weak. This is the case in GRO J0422+32 for which no
S-wave was present in H$\alpha$ spectroscopy during quiescence
(Harlaftis et al. 1999).  See also Mason et al. (2001) and Levitan et
al. (2011) for examples of lack of/weak S-waves in a 1.8 hr dwarf nova
and in an AM CVn system, respectively.  Second, the average radial
velocity offset found during quiescence is fully consistent with the
$\gamma$ velocity of $\sim-150 $\,km\,s$^{-1}$ that can be derived from the diagnostic diagram for this Balmer line
during outburst (fig. S2 in Corral-Santana et al. 2013). Third, Corral-Santana et
al. found $\gamma$ and a periodic (likely orbital) modulation from
spectroscopic data obtained in two outburst epochs separated by
$\sim20$ days. The detection of an orbital modulation in the combined
data sets is possible if any additional velocity shifts in the line
centroid on a timescale longer than the orbital period were small or
if they had similar magnitude and sign at the time of the
observations.  For CVs and LMXBs with $q \lesssim 1/4$, long-term line shifts
are expected during outburst due to disc eccentricity and precession.
The formation of such a disc is attributed to the tidal influence of
the donor star over a disc that during outburst radially expands to
reaches the 3:1 resonance radius (see e.g. Hirose \& Osaki 1990,
Whitehurst \& King 1991).  The variations in the emission line
centroids observed in XTE J1118+480 (${P_{orb}} = 4.08$ hr,
q=0.037; Zurita et al. 2002, Torres et al. 2004) and AM Canum Venaticorum ($P_{orb} = 17$ min,
$q=0.18$; Roelofs et al. 2006) provide evidence for the existence of large
amplitude velocity shifts driven by a precessing disc.  When present
they can make it difficult to recover any orbital modulation
superimposed on the lines by using data sets spanning many orbits and
lacking enough orbital phase coverage at each epoch of observation.
It appears therefore that long-term changes in the line centroid due
to a precessing accretion disc were small or absent in J1357.2 (at
least) at the time of the spectroscopic observations.  

In support of the possibility of a non-precessing disc on J1357.2 is
the absence of superhumps in the outburst light curves presented in
Corral-Santana et al. (2013).  Superhumps are photometric modulations
that are driven by outbursts and characterized by having a periodicity
a few percent longer than the actual orbital period. This photometric
behaviour is explained as due to the presence of the elliptical and
precessing disc. Superhumps have been found during the outburst
decline of the short orbital period black hole LMXBs GRO J0422+32
(O'Donoghue \& Charles 1996), XTE J1118+480 (Uemura et al. 2000,
Zurita et al. 2002, 2006) and Swift J1753.5-0127 (${P_{orb}} = 3.2$
hr, Zurita et al. 2008). The apparent lack of superhumps in J1357.2 is
striking, but not impossible given that these features do not
necessarily develop or persist during all outbursts. A clear example is the
eclipsing AM CVn system SDSS J0926+3624 (${P_{orb}} = 28$ min,
$q=0.04$) which exhibited superhumps in an outburst in 2006 and lacked
these features in an outburst with similar brightness amplitude that
occurred in 2009 (Copperwheat et al. 2011). Alternatively, the
considerable dipping in the outburst light curve of J1357.2 might not
have permitted the detection of the superhump signal which has been
argued to be of  low amplitude for high inclination LMXBs (Haswell et
al. 2011).  However, see   Mason et al. (2008) and  Hakala et
al. (2008) for the detection of  superhumps in the  ADC neutron  star
system MS 1603.6+2600 (${P_{orb}} = 1.9$ hr).  

In summary, the absence of S-wave components in the H$\alpha$ line
during quiescence, the similar values of the averaged line centroid
during quiescence and the $\gamma$ measured during outburst together
with the apparent lack during outburst of significant long-term
changes in the line strongly support the claim that the averaged
centroid of the H$\alpha$ line closely represents the systemic radial
velocity of J1357.2.  Since our radial velocities could have been
affected to a greater or lesser degree by systematic effects (Section
2), we prefer and adopt for the rest of the discussion the value of
$\gamma \sim -150$ km s$^{-1}$ derived in Corral-Santana et
al. (2013).

\subsection{The space velocity}

Since the distance to J1357.2 is poorly constrained (0.5--6.3\,kpc) and 
the proper motion is unknown, we cannot draw definitive conclusions on 
the space velocity of the source.  Given the high Galactic latitude of 
J1357.2 ($b=50^{\circ}$) the range of possible distances implies that 
the source lies between 0.4 and 4.8\,kpc above the Galactic Plane.  With 
the known position and systemic radial velocity ($-150$\,km\,s$^{-1}$), 
we can compute the Galactic space velocity components $U$, $V$ and $W$ 
for a grid of possible proper motions (in both right ascension and 
declination) and distances, using the transformations of Johnson \& 
Soderblom (1987).  We find that the $W$ velocity component (motion 
perpendicular to the Galactic plane) is directed back towards the Plane, 
unless the proper motions are large enough that the velocity components 
parallel to the plane of the disc deviate significantly (by 
$>150$\,km\,s$^{-1}$) from the standard Galactic rotation at 
240\,km\,s$^{-1}$ (Reid et al.\ 2014).  This would imply that the source 
was originally at an even larger distance from the Galactic Plane.  
Either it is a halo object, or (more likely) was formed in the disc and 
subsequently launched into an extremely elliptical orbit by a 
significant natal kick during the supernova explosion in which the black 
hole was formed, as also inferred to have occurred in GRO J1655-40 
(Brandt, Podsiadlowski \& Sigurdsson 1995) and XTE J1118+480 (Mirabel et 
al.\ 2001).

To place better constraints on the space velocity of J1357.2 would 
require knowledge of the source proper motion.  However, since the 
quiescent counterpart of J1357.2 is fainter than Gaia's limiting 
magnitude of $G=20$, an accurate proper motion for J1357.2 would require 
VLBI observations, either of the (as-yet undetected) quiescent radio 
emission, or of future transient outbursts (e.g.\ Mirabel et al. 2001; 
Russell et al., 2015).
 
\subsection{The orbital inclination}

An edge-on nature for J1357.2, as proposed to explain the profound
dips in the optical light curves (Corral-Santana et al. 2013) is
questionable.  As argued by Armas-Padilla et al. (2014a) the X-ray
spectrum during outburst shows an excess of soft  emission attributable to
non-occulted inner regions  of the accretion disc. Moreover the X-ray
spectrum lacks emission/absorption X-ray features frequently observed
in ADC sources (see sec. 4.1 in Armas-Padilla et al. 2014a) or in
high-inclination  neutron star LMXBs (Boirin et al.~2005;
D{\'{\i}}az Trigo \& Boirin 2013).
Furthermore, there is another significant difference between J1357.2
in outburst and X-ray bright high-inclination systems (not necessarily
ADC sources). In addition to extinction from interstellar matter, the
$N_H$ as measured in X-ray spectral fits usually includes the
absorption associated to matter local to the source. The latter can be
the dominant contribution to $N_H$ when reddening towards the source
is very low.  In several high-inclination systems the $N_H$ determined
from X-ray spectral fits is indeed dominated by extinction local to
the source. For instance, ultraviolet observations  have enabled reliable 
determinations of the interstellar reddening E(B-V) towards EXO
0748-676 and X2127+119 (which have ${P_{orb}} = 3.8$ hr and ${P_{orb}}
= 17.1$ hr, respectively): for both sources $E(B-V) \sim 0.06$ mag
(Ioannou et al. 2003, Pearson et al. 2006) equivalent to ${N_H} \sim 3
\times 10^{20}$ cm$^ {-2}$ when using Bohlin et al. (1978)'s $E(B-V)$ to
$N_H$ scaling. These values are much lower than $N_H$ values found
from X-ray fitting (see EXO 0748-676 Bonnet-Bidaud et al. 2001 and
White \& Angelini 2001 for X2127+119).

In other high-inclination sources the $N_H$  found for
dipping and/or eclipsing LMXBs when using the  out-of-dips or eclipse
X-ray emission is typically  of the order of $10^{21}$ cm$^{-2}$ as
found for   MS 1603.6+2600 (${P_{orb}} = 1.9$ hr, Hakala et al. 2005),
XTE J1710-281 (${P_{orb}} = 3.8 $ hr (Young et al. 2009), X1822-371
(${P_{orb}} =5.6$ hr, Somero et al. 2012, Iaria et al. 2013),  and
other longer/shorter orbital-period high-inclination neutron star
systems (D{\'{\i}}az Trigo et al. 2006).

Both the absolute value of $N_H$ and the difference between the $N_H$
from X-ray spectral fits and the optically-derived equivalent $N_H$ are
very low for J1357.2. {\it Swift} data delivered $N_H$ of $(1.2 \pm
0.7) \times 10^{20}$ cm$^{-2}$ while fits to XMM data could not
constrain it (Krimm et al.  2011b, Armas-Padilla et al. 2013).  We
can use the EWs of the interstellar Na doublet found in outburst
(Torres et al.~2011) to estimate the interstellar reddening towards
J1357.2.  The resolved Na D1 and D2 components, have $\sim0.3$ \AA~and
$\sim0.2$ \AA, respectively. We derive $E(B-V) \sim 0.1$ from the Na
D1 line EW using the calibration of Munari \& Zwitter (1997) and 0.05
from the Na D2 EW assuming a D2/D1 ratio of 2 (optically thin limit).
Thus E(B-V) is likely $0.05 - 0.1$ and thereby ${N_H} = (3- 6) \times
10^{20}$ cm$^{–2}$ which is very similar to the $N_H$ derived from the
X-ray spectral fits given the systematic effects in both methods employed. On the basis of this, we conclude that J1357.2 has a low $N_H$ when
compared to the value expected from a dipping and/or eclipsing high
inclination LMXB.  Possibly, the scale height of the material
responsible for the local $N_H$ enhancement is lower in J1357.2, or
alternatively, the system inclination is lower than that of the high
inclination systems we compared with. But this seems to suggest that
the inclination in J1357.2 is lower than what has been suggested
previously and probably $\lesssim 80^{\circ}$. This would explain the
lack of disc eclipses in the continuum or H$\alpha$
lines. Nevertheless, the orbital inclination of J1357.2 cannot be too
low given the observed emission line characteristics.  In particular,
the depth of the absorption core defined by the line peaks is similar
to that observed during quiescence in eclipsing CVs (Marsh \& Horne
1987) and is a strong indicator that the system has a high orbital
inclination (Horne \& Marsh 1986). We conclude that the inclination of
J1357.2 is probably between 70--80$^\circ$. 

\section{Conclusions}

Optical spectra of J1357.2 during quiescence were analyzed in this
work. The average data shows a continuum lacking photospheric lines
from the donor star.  Thus we spectroscopically prove that synchrotron
emission from a jet and/or thermal emission from the accretion disc
fully veil the light contribution from the donor star at optical
wavelengths.  Broad strong H$\alpha$ and weaker He{\sc i} emission
lines are present in the data. The  time-resolved H$\alpha$ line
profile shows no strong S-wave patterns or other obvious periodic
behavior such as Z-wave caused by eclipses. We also find that no large
long-term radial velocity shifts were  present, either at the time of
our spectroscopy, or of the spectroscopy acquired in outburst.  Thus
the data supports a systemic velocity $\gamma = -150$ km s$^{-1}$ on
the basis of the average radial velocity properties. From the
H$\alpha$ peak-to-peak separation we constrain ${K_2}$ to be $\gtrsim
796 \pm 7$ km s$^{-1}$ and $q \lesssim 0.054$. We estimate an
interstellar  absorption column  towards the source of $(3- 6) \times
10^{20}$ cm$^{-2}$ which is comparable to that derived from X-ray
fitting and lower than expected from an edge-on source. Therefore the
low X-ray brightness and spectral shape during outburst  are unlikely
due to a geometric effect (accretion disc corona system) and they are
consistent with the source being in the low-hard state during the
entire outburst.  Although J1357.2 appears not to be one of the many
missing eclipsing black-hole LMXBs, the system has high inclination as
inferred from the emission line morphology.  Finally,  we infer that
the present Galactic kinematics of J1357.2 are such that its $W$  space velocity component  
is directed towards the Galactic Plane unless the proper motion is
  substantial. 

\section*{Acknowledgments} \noindent 

We thank the anonymous referee for useful comments on the manuscript.
We would like to thank Remco de Kok for an independent search for the cross-correlation sinal from the donor star in J1357.2
and  Jorge Casares for providing us with spectral templates used in these analysis. JCAMJ is the recipient of an Australian Research Council (ARC) Future Fellowship (FT140101082), and also acknowledges support from an ARC Discovery Grant (DP120102393). DS acknowledges support from STFC
through an Advanced Fellowship (PP/D005914/1) as well as grant
ST/I001719/1

\newpage

\begin{figure*}
\includegraphics[width=6.0in, angle=90.0]{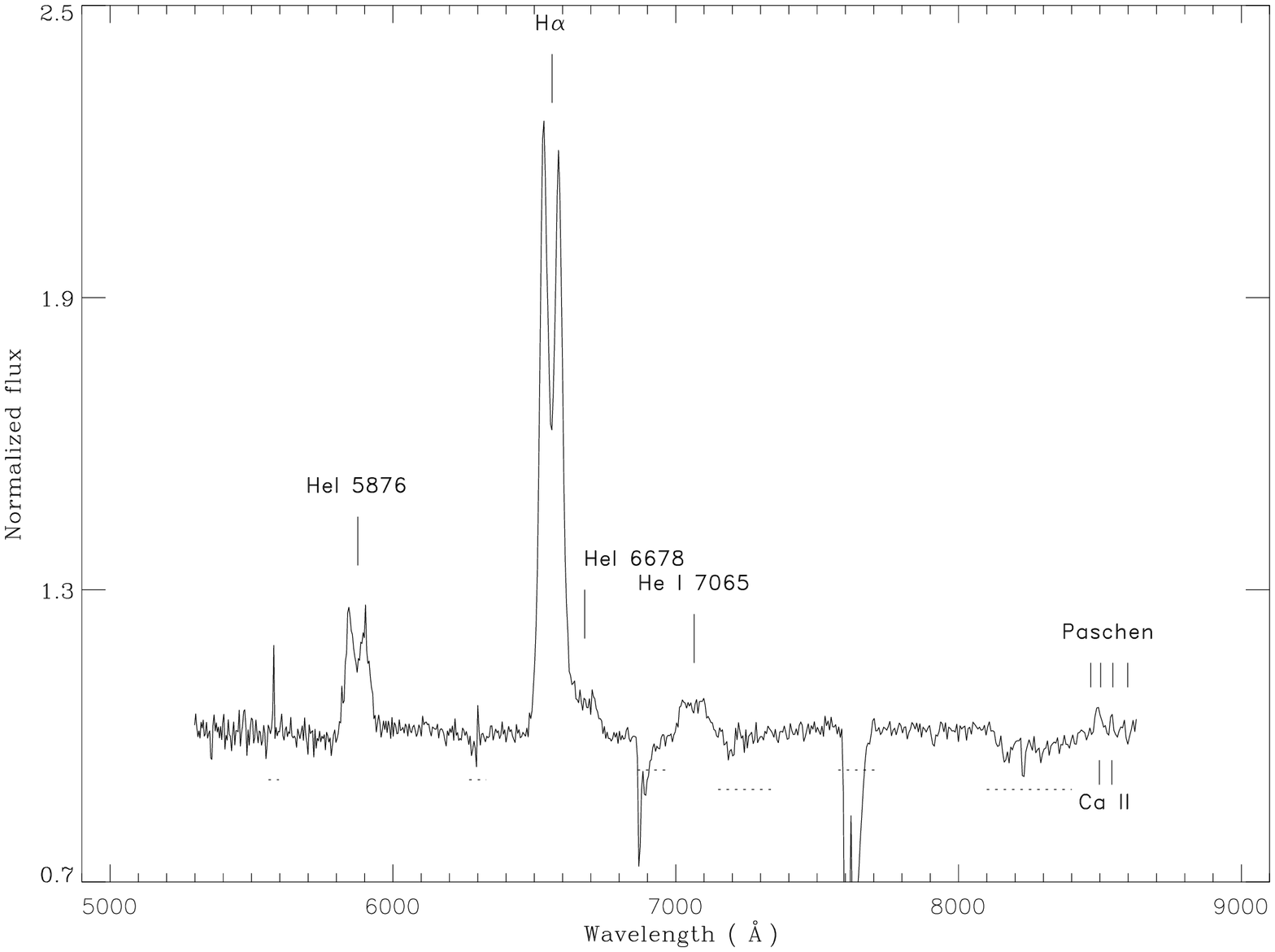}
\caption{Continuum-normalized and averaged optical spectrum of J1357.2
  in quiescence.  The rest wavelengths for H$\alpha$ and identified
  He{\sc i} disc line are provided.  For the sake of clarity, we also
  mark the rest wavelengths for the two Ca{\sc ii} infrared components
  at $\lambda\lambda8498,8542$ and Paschen lines at
  $\lambda\lambda8467,8502,8545,8598$. The dashed lines below the
  source continuum mark the location of strong telluric features and residual sky emission lines.}
\label{licus}
\end{figure*}

\newpage

\begin{figure*}
\includegraphics[width=6.0in, angle=90.0]{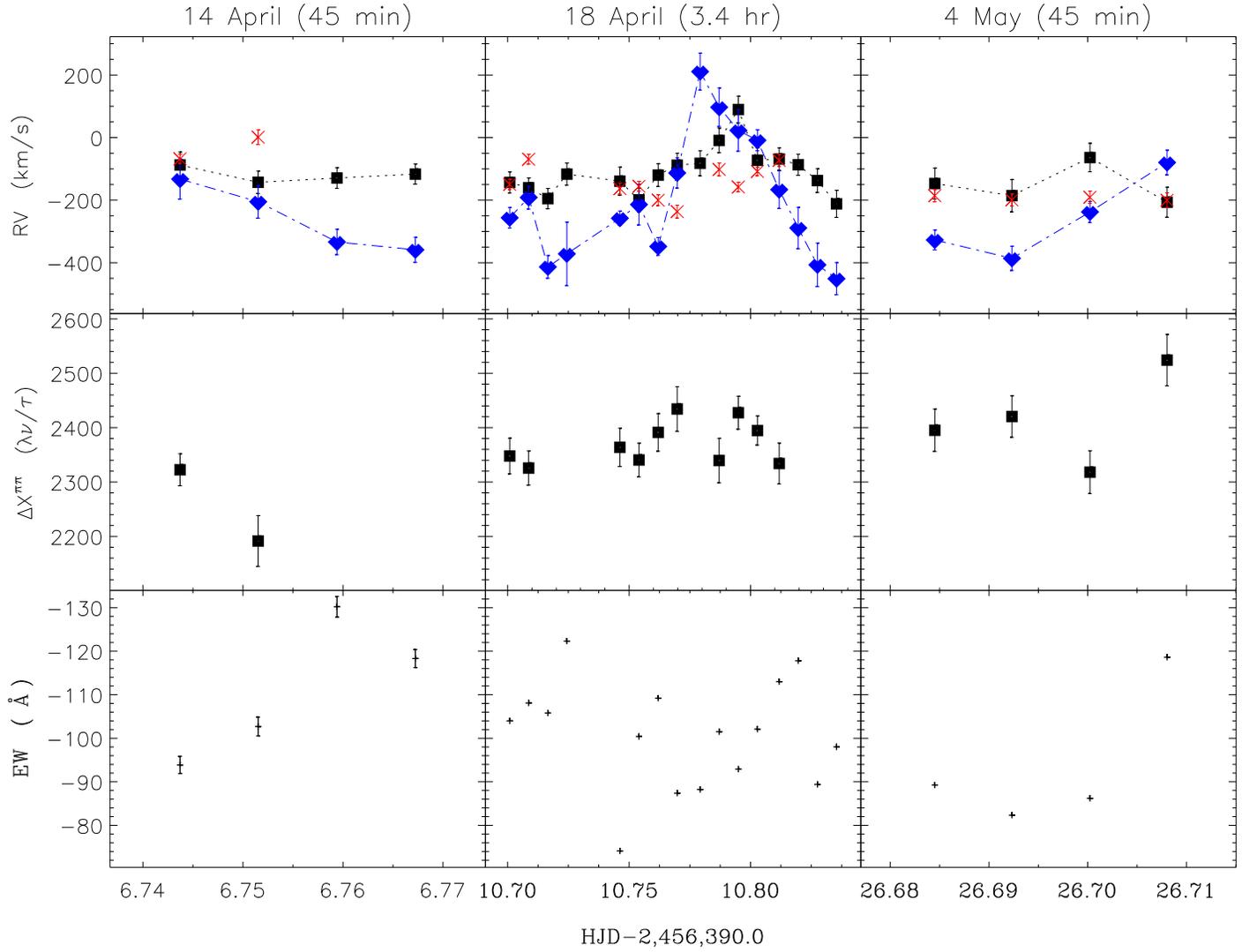}
\caption{H$\alpha$ emission line variability over the three nights of observations. Top: radial velocities from a single Gausian fit ($\blacksquare$) and the double-Gaussian technique (\protect\rotatebox[origin=c]{45}{$\blacksquare$}). The line centroid derived with a 2-Gaussian model is also shown  (X). Middle: peak-to-peak separation ($\Delta V^{pp}$) from the 2-Gaussian model. Bottom: EW measured in the interval $\lambda\lambda6460-6650$.}
\label{licus}
\end{figure*}

\begin{figure*}
\includegraphics[width=5.5in, angle=90.0]{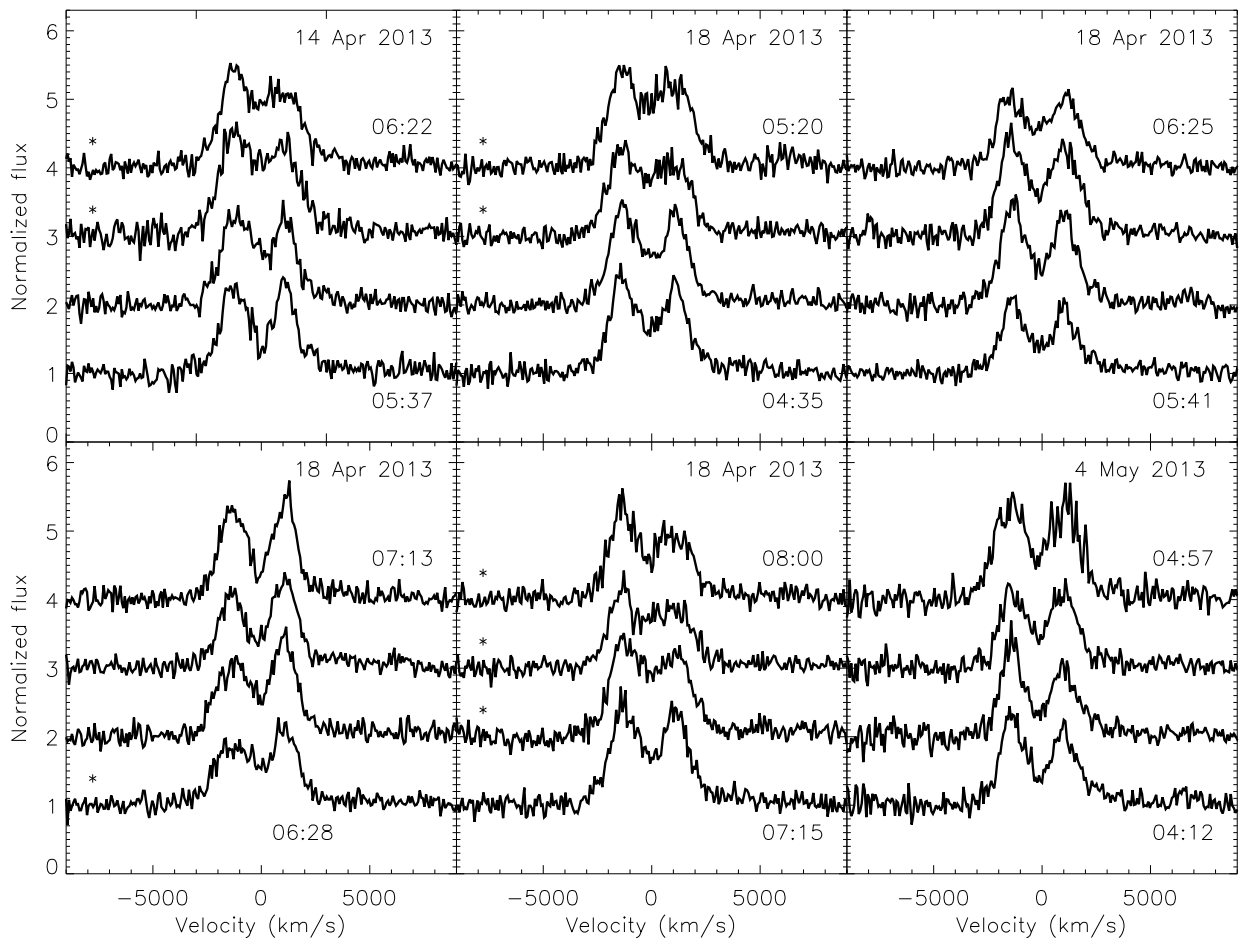}
\caption{The 24 individual H$\alpha$ line profiles covering different observing times. For the sake of clarity each consecutive spectrum has been offset in the Y-direction. The starting and ending UT times of the spectroscopy obtained in the same observing block are given in each panel. Note the changes in the depth of the abosorption component in the line center. An * marks the eight asymmetric profiles discussed in Section 3.2.} 
\label{licus}
\end{figure*}

\begin{table*}
  \caption{Emission line parameters: Centroid and peak-to-peak
    separation ($\Delta V^{pp}$) derived from a 2-Gaussian fit to the line
    profiles (except for He{\sc i} $\lambda7065$ where the line was fit with a
    single Gaussian). The FWHM was derived from a single Gaussian fit and 
corrected for the instrumental spectral resolution.  The uncertainties for the centroid,
    peak-to-peak separation and FWHM were calculated after scaling the
    data error bars to yield a fit with $\chi{^2}/d.o.f = 1$. The
    uncertainties in the EWs are estimated by determining
    the scatter in the values derived 
    when selecting different wavelength intervals to set the local
    continuum level. The FWZIs are given as lower limits since
the ability to determinate the extension of the line wings is usually set by the signal-to-noise in the data.}

\label{log}
\begin{center}
\begin{tabular}{lccc}
\hline
                            &  H$\alpha$  & He{\sc i} 5876 &  He{\sc i} 7065    \\       
\hline
                             &                                &                             &                              \\                
Centroid (km/s)  & $ -137 \pm 8 $      &  $-180 \pm 40$  &  $-100 \pm 100$     \\        
$\Delta V^{pp}$ (km/s)  & $2340 \pm 20 $    & $2640 \pm  70 $ &  ---                       \\        
 FWHM (km/s)   & $4025 \pm 110$  &   $4500 \pm 200$  & $ 4290 \pm 250 $   \\       
 FWZI (km/s)     & $\gtrsim 6300$              &  $\gtrsim 7040 $           & $\gtrsim 5900  $            \\       
EW (\AA)          & $-92 \pm 1$    & $ -19 \pm 1$    &  $-9 \pm 2$     \\       
                             &                    &                                        &                              \\                
  \hline
\end{tabular}		      
\end{center}
\end{table*}

\end{document}